\documentclass[aps,prl,showpacs,superscriptaddress,twocolumn,final,numerical,10pt]{revtex4-1}
\usepackage{hyperref}
\usepackage{verbatim}
\usepackage{amsmath}
\usepackage{latexsym}
\usepackage{revsymb}
\usepackage{yfonts}
\usepackage{ifthen}
\usepackage{color}
\usepackage{natbib}
\usepackage{amsfonts}
\usepackage{amsmath}
\usepackage{amssymb}
\usepackage{amsthm}
\usepackage{graphicx}
\usepackage{bm}
\usepackage{bbm}
\usepackage{float}
\usepackage{lipsum}
\usepackage{capt-of}
\usepackage{framed}
\definecolor{shadecolor}{rgb}{0.92,0.92,0.92}

\begin{document}

\title{Secure Quantum Key Distribution}

\author{Hoi-Kwong Lo$^{\dagger}$}
\affiliation{Center for Quantum Information and Quantum Control, Dept. of Physics and Dept. of 
Electrical \& Computer Engineering, University of Toronto, M5S 3G4 Toronto, Canada}
\author{Marcos Curty$^{\dagger}$}
\affiliation{EI Telecomunicaci\'on, Dept. of Signal Theory and Communications, University of Vigo, E-36310 Vigo, Spain}
\author{Kiyoshi Tamaki$^{\dagger}$}
\affiliation{NTT Basic Research Laboratories, NTT Corporation, 3-1, Morinosato-Wakamiya Atsugi-Shi,
243-0198, Japan \\ 
\noindent $^{\dagger}$ All authors contributed equally to this work}

\date{\today}

\begin{abstract}
Secure communication plays a crucial role in the Internet Age. Quantum mechanics may revolutionise cryptography as we know it today. 
In this Review Article, we introduce the motivation and the current state of the art of research in quantum cryptography. In particular, we 
discuss the present security model together with its assumptions, strengths and weaknesses. After a brief introduction to recent
experimental progress and challenges, we survey the latest developments in 
quantum hacking and counter-measures against it.
\end{abstract}

\maketitle

\noindent With the rise of the Internet, the importance of cryptography is growing every day. 
Each time we make an on-line purchase with our 
credit cards, or we 
conduct financial transactions using Internet banking,
we should 
be concerned with secure communication. Unfortunately, the security of conventional cryptography is often based on computational assumptions. For instance, 
the security of the RSA scheme~\cite{rsa}, the most widely used public-key encryption scheme, is based on the presumed hardness of factoring. Consequently, conventional 
cryptography is vulnerable to unanticipated advances in hardware and algorithms, as well as to quantum code-breaking such as Shor's efficient algorithm~\cite{shor}
for factoring. Government and trade secrets are kept for decades. An eavesdropper, Eve, may simply save communications sent in 
2014 and wait for 
technological advances. If she is able to factorise large integers in say 2100, she could retroactively break the security of data sent 
in 2014.

In contrast, quantum key distribution (QKD), the best-known application of quantum cryptography, 
promises to achieve the Holy Grail of cryptography---unconditional security in communication. By unconditional security 
or, more precisely, $\epsilon$-security, as it will be explained shortly (see section discussing the security model of QKD), Eve is not restricted by computational assumptions
but she is only limited
by the laws of physics.
QKD is a remarkable solution to long-term security since, in principle, it offers security for eternity. Unlike conventional cryptography, which 
allows Eve to store a classical transcript of communications, in QKD, once a quantum transmission is done, there is no classical 
transcript for Eve to store. See Box 1 for background information on secure communication and QKD.
\begin{widetext}
\begin{shaded}
\noindent{\bf Box 1 $|$ Secure communication and QKD.}

\noindent{\bf Secure Communication: }
Suppose a sender, Alice, would like to send a secret message to a receiver, Bob, through an open communication channel. Encryption is needed. If they share a common string of secret bits, called a key, Alice can use her key to transform a plain-text into a cipher-text, which is unintelligible to Eve. In contrast, Bob, with his key, can decrypt the cipher-text and recover the plain-text. In cryptography, the security of a crypto-system should rely solely on the secrecy of the key. The question is: how to distribute a key securely? In conventional cryptography, this is often done by trusted couriers.  Unfortunately, in classical physics, couriers may be brided or compromised without the users noticing it. 
This motivates the development of quantum key distribution (QKD).

\noindent{\bf Quantum Key Distribution: } 
The best-known QKD protocol (BB84) was published by Bennett and Brassard in 1984~\cite{bb84}. Alice sends Bob a sequence of photons prepared in different polarisation states, which are chosen at random from two conjugate bases. For each photon, Bob selects randomly one of the two conjugate bases and performs a measurement. He records the outcome of his measurement and the basis choice. Through an authenticated channel, Alice and Bob broadcast their measurement bases. They discard all polarisation data sent and received in different bases and use the remaining data to generate a sifted key. To test for tampering they compute the quantum bit error rate (QBER) of a 
randomly selected subset of data and verify that the QBER is below a certain threshold value. By applying classical post-processing protocols such as error correction and privacy amplification, they generate a secure key. This key can be used to make the communication unconditionally secure by using a one-time-pad protocol~\cite{vernam}.

\noindent{\bf One-time-pad protocol: }
The message is represented by a binary string. The key is also a binary string of the same length as the message. For encryption, a bitwise exclusive-OR (XOR) is performed between the corresponding bits of the message and the key to generate a cipher-text. Decryption is done by performing a bitwise XOR between the corresponding bits of the cipher-text and the key. For one-time-pad to be secure, the key should not be re-used. 
\end{shaded}
\end{widetext}
\begin{figure*}
\begin{center}
 \includegraphics[scale=0.394,angle=0]{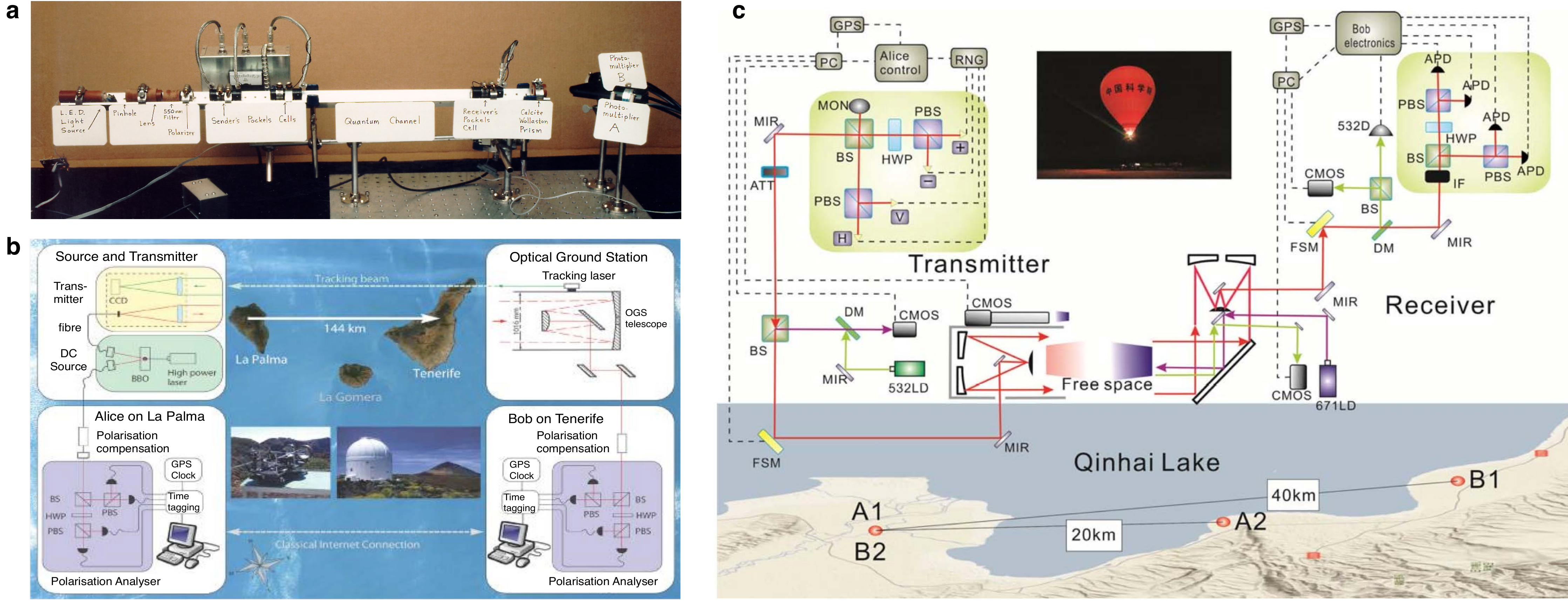}
 \end{center}
\caption{
{\bf $|$ Progress in free-space QKD implementations.} {\bf a}, 
First free-space demonstration of QKD~\cite{first} realised two decades ago over a distance of $32$cm. The 
system uses a light emitting diode (LED) in combination with Pockels Cells to prepare and measure the 
different signal states. {\bf b}, Entanglement-based QKD setup connecting the Canary Islands of La Palma and Tenerife~\cite{free_QKD}. The length 
of the optical link is $144$km. {\bf c}, 
Schematic diagram of a decoy-state BB84 QKD experiment between ground and a hot-air balloon~\cite{sat2}. 
This demonstration can be seen as a first step 
towards QKD between ground and Low Earth Orbit Satellites. In the figure: 
(HWP) half-wave plate, (MON) monitor window, (MIR) mirror, (ATT) attenuator, (DM) dichroic mirror, (532LD) $532$nm laser,
(FSM) fast steering mirror, (671LD) $671$nm laser, (532D) $532$nm detector, and (IF) interference filter. Figures from: {\bf a}, Ref.~\cite{first}; {\bf b}, Ref.~\cite{free_QKD}; and {\bf c}, Ref.~\cite{sat2}.
\label{fig1}}
\end{figure*} 
\noindent {\bf Achievements and future goals in QKD.}
On the theory side, a landmark accomplishment has been rigorous security proofs of QKD protocols. 
Recently, a ``composable'' definition~\cite{comp1,comp2} of the security of QKD 
has been obtained. Stable QKD over long distances (of 
the order of $100$km) has been achieved in both fibres and free-space~\cite{stableQKD,free_QKD}. 
Commercial QKD systems are currently available in the market.
Field test demonstrations 
of QKD networks have been 
done~\cite{net1,net2,swissQ,China_net1,China_net2,ChineseQKD,demQKD,net3}. High detection efficiency single-photon detectors at telecom 
wavelengths have been developed~\cite{det1,High-efficient SNSPD21,High-efficient SNSPD3,restelli13}.
In short, 
QKD is 
already mature enough for real-life applications.
As an illustration, Fig.~\ref{fig1} shows the tremendous progress
that has been made in free-space QKD over the last two decades. It compares 
the first lab demonstration performed in $1992$~\cite{first} with two 
recent QKD implementations that connect, respectively, two Canary Islands~\cite{free_QKD} and a ground station 
with a hot-air balloon~\cite{sat2}.

And, what researchers are aiming to do now?
As will be discussed in the rest of the paper, to guarantee unconditional security in actual QKD implementations, 
researchers are working hard 
to bridge the gap between theory and practice. Also, the development of high-speed QKD systems, together 
with the ability of 
multiplexing 
strong classical signals with weak quantum 
signals in the same optical fibre, for example, via wavelength division multiplexing (WDM), are 
major research challenges of the field. Moreover, 
researchers are studying QKD network set-ups with both trusted and untrusted nodes. 
The feasibility of ground to satellite QKD has also 
attracted a lot of research attention~\cite{sat,sat2}.

\quad

\noindent {\large\bf Security model of QKD} 

\noindent 
Intuitively speaking, the security of 
QKD 
is measured with 
respect to a perfect key distribution scheme
where Alice and Bob 
share a true random secret key. 
More precisely, we say that a QKD system is $\epsilon$-secure if and only if 
the probability distribution of an outcome of {\it any} measurement performed on the QKD scheme and the resulting 
key deviates at most $\epsilon$ from the one of the perfect key distribution protocol and 
the perfect key~\cite{comp1, comp2,sec_prac}. 
A typical value for $\epsilon$
is $10^{-10}$. 
However, in principle  Alice and Bob could select $\epsilon$ as small as they want, just by applying enough privacy 
amplification. 

Of course, since a secret key is a resource for other cryptographic 
protocols ({\it e.g.}, the one-time-pad method), it is not enough to consider the security of the QKD protocol alone. Instead, one has 
to evaluate the security of the generated key when it is employed in a crypto-system. 
This notion is known as ``composable'' security. 
Fortunately, QKD is composably secure~\cite{comp1, comp2,sec_prac}. That is, 
if we have a set of cryptographic protocols (which may include QKD), 
each of them having a security parameter $\epsilon_i$, as part of a certain cryptographic scheme, then the security of the whole 
system is given by $\sum_{i}\epsilon_i$. 

\quad

\noindent {\bf Progress in security proofs.}
Once we have presented the security definition of QKD, next we discuss the security of a particular QKD implementation: the BB84 scheme~\cite{bb84}.
In its original theoretical proposal, Alice sends Bob single-photon states. 
However, as practical and efficient single-photon sources are yet to be realised,
most implementations of the BB84 protocol are based on
phase-randomised weak coherent state pulses (WCPs) with a typical average photon number of $0.1$ or higher.
These states can be easily prepared using standard semiconductor lasers and calibrated attenuators.
The main drawback of these systems, however, arises from the fact that some signals may contain more than one photon prepared in the same quantum state.
If Eve performs, for instance, the so-called Photon-Number-Splitting (PNS) attack \cite{PNS} on the multi-photon pulses,  
she could obtain full information about the part of the key generated with them
without causing any noticeable disturbance. That is, in BB84 only the single-photon states sent by Alice and detected by Bob can 
provide a secure key. 
Fortunately, to distill a key from these single-photon contributions it is enough if Alice and Bob can estimate a lower bound for the total number of such events, {\it i.e.}, they 
do not need to identify which particular detected pulses originate from single-photon emissions~\cite{GLLP}.
In the case of the BB84 scheme, this estimation procedure 
must assume the worst case scenario where Eve blocks as many single-photon pulses as possible. 
As a result, it turns out that its key generation rate scales as $\eta^2$, where $\eta$ denotes the transmittance of the quantum channel.
This quantity has the form $\eta=10^{-\frac{\alpha d}{10}}$, where $\alpha$ is the loss coefficient of the channel 
measured in dB/km ($\alpha\approx0.2$ dB/km for standard commercial fibres) and $d$ is the covered distance measured in km. 

In reality, however, Eve may not be monitoring the quantum channel and performing a PNS attack. 
To improve the achievable secret key rate in general, therefore, 
is necessary 
to estimate more precisely the amount of single-photon pulses detected by Bob.
This can be done using the so-called decoy-state method~\cite{Decoy1,Decoy2,Decoy3,Decoy4,decoyexp1,decoyexp2,decoyexp3,decoyexp4,decoyexp5,decoyexp6}, 
which can basically reach the performance of single-photon sources,
where the key generation rate scales linearly with $\eta$. The procedure is as follows. 
Instead of sending signals of equal intensity, Alice chooses first the intensity for each signal at random from a set of prescribed values. 
States sent in one
particular intensity are called signal states, whereas the states sent
with other intensities are called decoy-states. Once Bob has detected all the signals, Alice 
broadcasts the intensity used for each pulse. 
A crucial assumption here is that all other possible degrees of freedom of the signals (apart from the intensity) are 
equal for all of them. This way, 
even if Eve knows the total number 
of photons contained in a given pulse, her decision on whether or not send that signal to Bob cannot depend
on its intensity. That is, 
Eve's decision is based on what is known {\it a priori}. 
Consequently, 
the probability of having a detection event given that Alice sent a single-photon pulse is the same for the signal and decoy pulses. As a result, 
Alice and Bob can estimate the fraction of detected events that arise from single-photons more precisely. 
This technique is rather general and also very useful for other quantum cryptographic protocols~\cite{Decoy-application}. 

\quad

\noindent {\large\bf Experimental implementations} 
\begin{figure*}
\begin{center}
 \includegraphics[scale=0.348,angle=0]{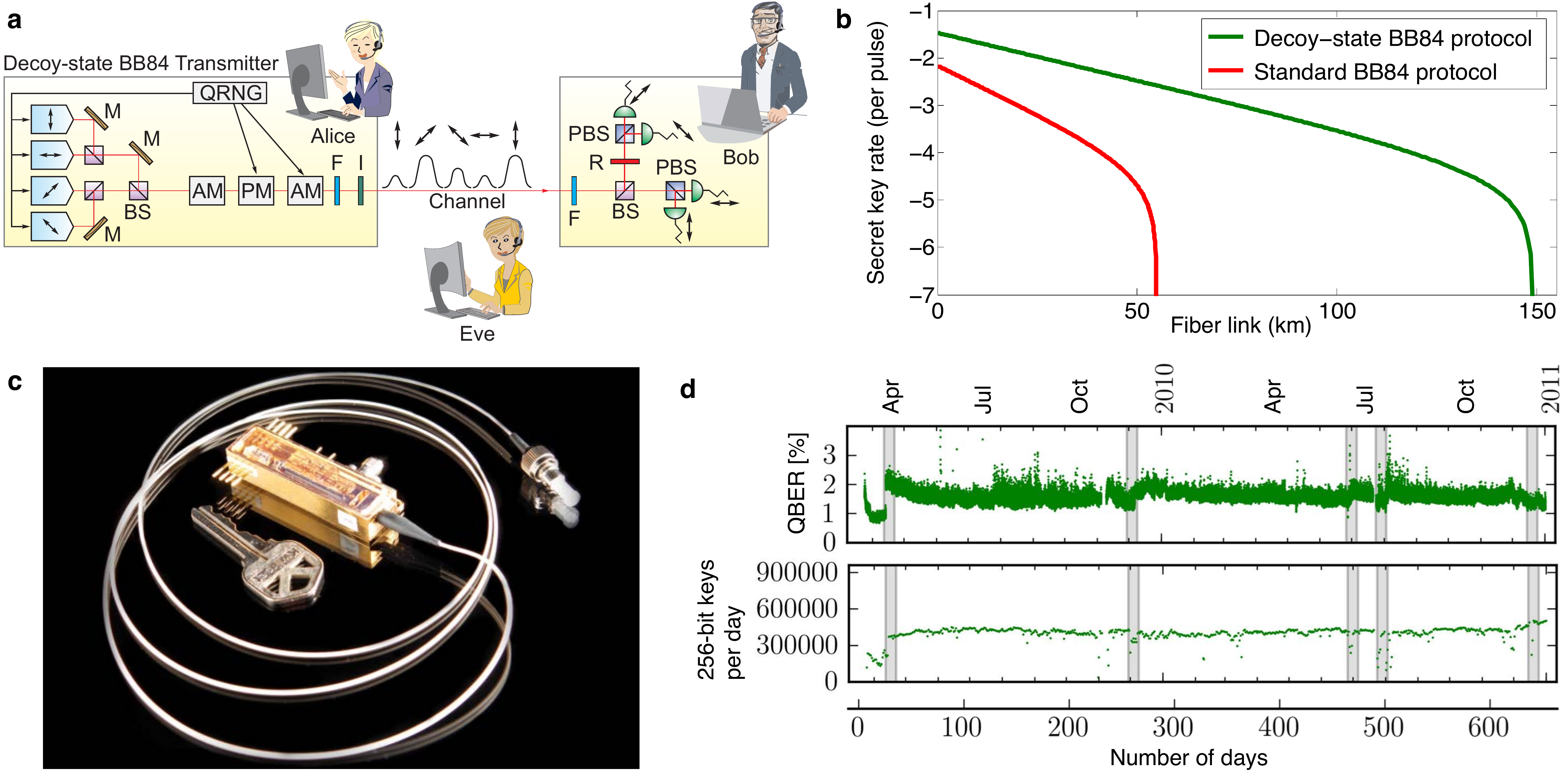}
 \end{center}
 \caption{
 {\bf $|$ Experimental QKD.} {\bf a}, Schematic diagram of the decoy-state BB84 protocol~\cite{Decoy1,Decoy2, Decoy3,Decoy4,decoyexp1,decoyexp2,decoyexp3,decoyexp4,decoyexp5,decoyexp6}
 based on polarisation coding. Four lasers are used to prepare the
polarisations needed in BB84. Decoy-states
are generated with an intensity modulator (AM). 
On Bob's side,   
a $50:50$ beam-splitter (BS) is used to passively ensure a random measurement basis choice. Active receivers are also common.
In the figure: (PM) phase modulator, (M) mirror, (F) optical filter, (I) optical isolator, (R) polarisation rotator, and (PBS)  
polarising beam-splitter.
{\bf b}, Lower bound on the secret key rate (per pulse) in logarithmic scale for a 
BB84 set-up with two decoys (green line)~\cite{Decoy4}. 
In the short distance regime, the key rate scales linearly with the transmittance, $\eta$. 
(Red line) standard BB84 protocol without decoy-states~\cite{bb84,GLLP}; its key rate 
scales as $\eta^2$.
{\bf c}, Photo of a fibre-coupled modularly-integrated decoy-state BB84 transmitter based 
on polarisation coding~\cite{hughesQKarD}; it produces decoy-state BB84 signals 
at a repetition rate of 10 MHz.
{\bf d}, Performance of the SwissQuantum network~\cite{swissQ}.
This network run for more than one and a half years
in the Geneva metropolitan area, Switzerland. 
The data shown in the figure corresponds to a QKD link of $14.4$km; it highlights how stable current QKD set-ups are.
Figures from: {\bf c}, Ref.~\cite{hughesQKarD}; and {\bf d}, Ref.~\cite{swissQ}.
\label{scheme2}}
\end{figure*} 

\noindent Experimental realisations of 
QKD have made a huge progress in the last two decades. In practice, the signal transmission can be done through free-space 
(using a wavelength around $800$nm) or through optical fibres (using the second or third telecom windows, {\it i.e.}, wavelengths
around $1310$nm and $1550$nm, respectively). Also, current set-ups use different degrees of freedom to encode the relevant 
information into the optical pulses. As already introduced, an obvious choice for this is to employ the polarisation state of the photons. This technique, 
so-called polarisation coding, is mostly used in free-space QKD links. For optical fibre transmission, however, one usually selects 
other coding options such as, for example, phase-coding, time-bin coding or frequency coding. This is so
because polarisation in standard fibres is more susceptible to
disturbances due to birefringence and environmental effects.   

Fig.~2a shows conceptually how simple is the basic set-up for the decoy-state BB84 protocol 
when 
Alice and Bob use polarisation coding. 
The expected secret key rate (per pulse) as a function of the distance 
is illustrated in Fig.~2b. The cut-off point where the secret key rate
drops to zero depends on the parameters of the system (specially, on the channel transmission and 
on the efficiency and dark count rate of Bob's detectors),
and is typically around $150$-$200$km. 
For comparison, as shown in the Figure, (the corresponding lower bound on) the secret key rate for the standard BB84 
protocol {\it without} decoy-states is much lower.
Fig.~2c shows a photo of 
a fibre-coupled modularly-integrated decoy-state BB84 transmitter
developed by Los Alamos group~\cite{hughesQKarD}; 
it is similar in size to an electro-optic modulator.

Alice and Bob may enlarge further the covered distance by using entanglement-based QKD protocols~\cite{ekert91,ent0,ent2,ent3}, as these 
schemes  
can tolerate higher losses (up to about $70$~dB) than WCP-based protocols. For instance, they could employ a parametric down-conversion source
to generate polarisation entangled photons that are distributed between them. This source could be even controlled by Eve,
and it can be placed in the middle between the legitimate users. On the receiving side, both Alice and Bob measure the 
signals received using, for example, a BB84 receiver like the one shown in Fig.~2a. 
Two drawbacks of this approach are, however, that the systems are more involved than those based on 
WCPs, and
their secret key rate is usually lower in the low loss regime. Alternatively to polarisation coding, 
one can use as well, for instance,  
energy-time entangled photon pairs.  

For shorter distances (say below $100$km), there are other solutions that are simpler to implement experimentally. 
These are the so-called distributed-phase-reference QKD 
protocols~\cite{dpr1,dpr5,dpr9}. 
They differ from standard QKD schemes in that now Alice encodes the information 
coherently between adjacent pulses, rather than in individual pulses. 
This approach includes the differential-phase-shift (DPS)~\cite{dpr1,dpr5} and the coherent-one-way 
(COW)~\cite{dpr9} protocols. In the former, Alice prepares a train of WCPs of equal intensity and modulates their phases. 
On the receiving side, Bob uses a one-bit delay Mach-Zehnder interferometer, followed by two single-photon detectors, to measure the incoming pulses. 
Similarly, 
in the COW protocol all pulses share a common phase but now Alice varies their intensities. 

An important issue in any QKD implementation is its reliability and robustness in 
a real life environment. Fig.~2d shows the performance (as a function of time) of a QKD link from 
the SwissQuantum network 
installed in Geneva, Switzerland~\cite{swissQ}. It demonstrates the high stability of current QKD systems.

The protocols described above belong to the so-called discrete-variable QKD schemes. Another interesting option is 
to use continuous-variable systems (CV-QKD)~\cite{cv0,cv1,cv4}. The key feature of this solution is that now the detection device
consists of (homodyne or heterodyne) measurements of the light-field quadratures. Consequently, these protocols can be implemented 
with standard telecom components and do not require single-photon detectors, which makes them also very suitable for 
experimental realisations. 

\quad

\noindent {\bf QKD components and data-processing.}
For the optical layer of a QKD system, the following components are typically needed:

{\it 1. Light sources:} Attenuated laser pulses can be used as the signal source in QKD. It is standard to model the signal as a WCP. 
By applying a global phase randomisation, the state becomes a classical mixture of Fock states ({\it i.e.}, states of different photon numbers) 
with Poissonian 
distribution. 

{\it 2. Single-photon detectors:} 
Single-photon detection is the ultimate limit of the
detection of light. It is important not only in
QKD applications, but also in sensitive measurements in astronomy and bio-medical
physics.
Traditionally, two different types of detectors---Silicon detectors and InGaAs detectors---have been widely used in QKD. 
Silicon detectors are broadly employed for visible wavelengths ({\it e.g.}, 800nm) and in free-space implementations. They have rather high detection efficiency 
of around 50\%. InGaAs avalanche photodiodes (APDs) are often used in telecom 
wavelengths and in fibre optical communication. Previously, they suffered from 
low detection efficiency of around 15\%. Another drawback of InGaAs APDs was that they
had a rather long dead-time after a detection event,
which severely limited the detection repetition rate to only
a few MHz. 
In the last few years, however, new detector technologies
have been developed for QKD applications. This includes, for example: self-differencing APDs~\cite{det2,dixon09}, 
the sine wave gating technique~\cite{namekata,liang,namekata2}, a hybrid approach combining these 
last two methods~\cite{Zhang09}, as well as
superconducting 
nanowire single-photon detectors (SNSPDs)~\cite{det1}, which all can operate at 
GHz detection repetition rate. Also, the detection efficiency of InGaAs APDs has been improved to 
about 50\% at 1310nm~\cite{restelli13}, and
new types of
SNSPDs with a very high detection efficiency of around 
$93\%$ have been developed~\cite{det1,High-efficient SNSPD21,High-efficient SNSPD3}. The main drawback of these 
novel SNSPDs~\cite{det1,High-efficient SNSPD21,High-efficient SNSPD3}, however, 
is their operating temperature, which is at the moment 
of the order of $0.1$~K. 
The dark count rate of these high efficiency SNSPDs~\cite{det1,High-efficient SNSPD21,High-efficient SNSPD3},
of the order of 100 Hz, can be substantially improved by better
rejection of ambient photons using optical band-pass filters at the input
port of the SNSPDs~\cite{Shibata}.

{\it 3. Standard linear optical components:} polarising beam-splitters, beam-splitters, amplitude modulators 
and phase modulators are widely used in QKD applications.

{\it 4. Random number generators:} Random numbers are
needed for basis choice, bit value choice, phase-randomisation, 
intensity choice in the decoy-state method 
as well as
for data post-processing. High-speed random number 
generation is a key bottleneck in current QKD. Fortunately, there have been a lot of research activities in the subject. Quantum mechanics offers true randomness 
from the laws of physics~\cite{random1}. A simple way to build a quantum random number generator (QRNG) is to send a WCP through a $50:50$ 
beam-splitter and put two single-photon detectors on the two outgoing arms. The actual bit value (0 or 1) generated depends on which detector detects a photon. 
Other methods~\cite{random2b,random3,random3b} to design QRNGs such as using phase noise~\cite{random4} also exist.

{\it 5. Classical post-processing techniques:} Processes such as 
post-selection of data (typically called sifting), error correction and privacy amplification are used to correct errors in the quantum transmission 
and to remove any residual information that Eve might have on the raw key. The final result is a key shared by Alice and Bob that Eve almost surely has 
absolutely no information about. A bottleneck in high-speed QKD is the computational complexity of 
classical post-processing protocols, together with the processing of huge raw data in a very short time. 
Fortunately, progresses have been
made for algorithm speed-up using hardware-based ({\it e.g.}, FPGA) solutions~\cite{stableQKD}.

{\it 6. Authenticated channel:} For QKD to work, besides a quantum channel, Alice and Bob need to share an authenticated classical channel. Fortunately, only a 
rather short authentication key is needed for this. Such an authentication key may be provided in the initial shipment of the QKD system through a temper-resistant device.
Once a QKD session has succeeded, one can refurbish the authentication key from the key generated by QKD. In this sense, QKD is a key growing protocol.

If initially there is no shared key between Alice and Bob,
they may also use a classical solution for authentication
based on computational assumptions via a certifying authority, which
is a standard protocol in the Internet. Provided that such an authentication scheme is unbroken for a short period of time during the first QKD session, 
the first QKD session will be secure and will generate the subsequent authentication keys.

\quad

\begin{figure*}
\begin{center}
 \includegraphics[scale=0.4275,angle=0]{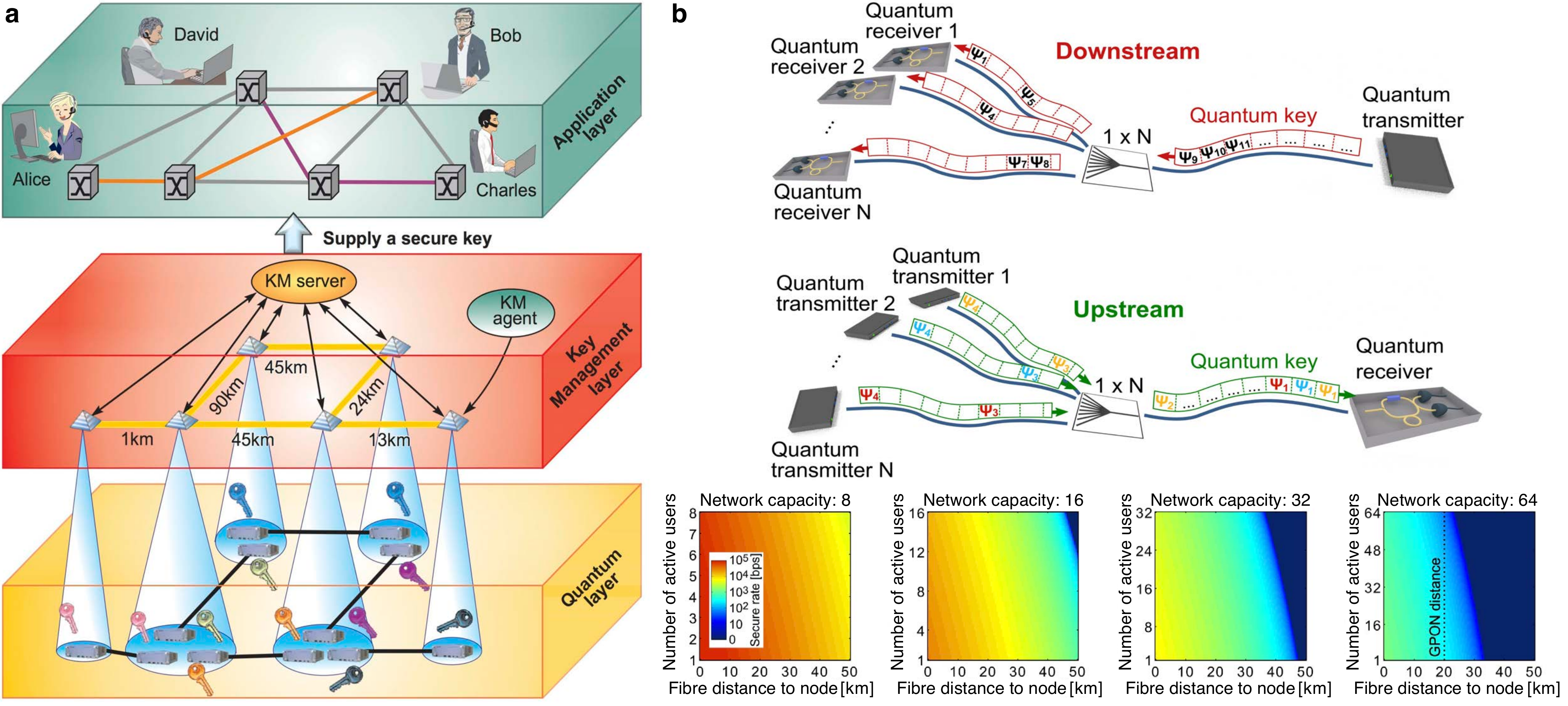}
 \end{center}
 \caption{
 {\bf $|$ QKD networks.} {\bf a}, 
Schematic representation of the layer structure of the Tokyo QKD network~\cite{demQKD}, which is based on a trusted node
architecture.
From the viewpoint of the users, the QKD layer and the key management layer can be treated as a black box that supplies them a secure 
key. On the application layer, they can enjoy applications such as secure video meetings and secure communication using 
smart phones. Different QKD networks have been implemented as well in other countries; see, for 
instance,~\cite{net1,net2,swissQ,China_net1,China_net2,ChineseQKD}.
{\bf b},
(Upper subfigure) Downstream versus upstream passive quantum access network. 
In the upstream approach~\cite{net3}, the single-photon detectors are located only at the network node.
This may reduce the costs of the network  
and allow a more efficient use of its detectors' bandwidth. (Lower subfigure) Estimated secret key rate 
per user for an upstream solution as a function of the distance and the number of active users in the network 
for various 
network capacities. Figures from: {\bf a}, adapted from Ref.~\cite{demQKD}; and {\bf b}, Ref.~\cite{net3}.
\label{fig3}}
\end{figure*} 
\noindent{\bf Industrial/application perspectives.}
The field of QKD attracts both fundamental research and industrial interests. 
As mentioned above, there are already commercial products  
that offer encryption solutions based on this technology.
Also, QKD networks have been recently deployed in 
USA~\cite{net1}, Austria~\cite{net2}, Switzerland~\cite{swissQ}, China~\cite{China_net1,China_net2,ChineseQKD} and 
Japan~\cite{demQKD}.
As an illustration, Fig.~3a shows the current structure of the Tokyo QKD network~\cite{demQKD}. It 
uses an
architecture based on trusted nodes, which are separated by distances that range between $1$km to $90$km. 
The network consists of three main layers: a QKD layer, a key management layer, and an application layer.
In the former, QKD systems that connect neighbouring nodes generate 
secret key material continuously, {\it i.e.}, without any maintenance~\cite{stableQKD}. 
This key, of the order of $300$~Kbps when the link loss is around 
$14.5$ dB~\cite{demQKD}, is forwarded to a key management agent placed 
in the key management layer. This agent monitors the key generation rate and the amount of stored key. 
Secure communication is possible between any nodes in the network by relaying on the secret key that is controlled by command of the 
key management server. 
From the viewpoint of the users, the QKD layer and the key 
management layer can be treated as a black box that supplies them a secure 
key. 
Such network could be employed, for instance, 
to provide secure communications with smart phones. Any time a user needs a fresh secret key
to protect his communication over the phone, he could connect to the QKD network 
and store in his device the key obtained for later use~\cite{demQKD}. 
Also, new architectures for QKD networks have been recently proposed
by the Toshiba group and by Los Alamos group. Fig.~3b compares the upstream passive quantum access network 
implemented by Toshiba~\cite{net3} with a downstream approach, whereas Fig.~2c is the compact transmitter prepared
by Los Alamos group~\cite{hughesQKarD}.

Lately, QKD systems have been used 
in the Swiss national elections
to 
protect the line that transmitted the ballots to the counting station. Also, they have secured a communication link 
at the 2010 FIFA World Cup competition in Durban, South Africa. 
Other potential applications of QKD include, for example, offsite backup, 
enterprise private networks, critical infrastructure protection, backbone protection and
high security access networks.

\quad

\noindent{\bf Technological challenges.}
As mentioned in the introduction,
researchers are working on designing and building high-speed
QKD systems~\cite{Townsend} and the ability of multiplexing
strong classical signals with weak quantum signals in the
same optical fibre~\cite{Chapuran2009, Toshiba2012,Toshiba2014}. Theorists are developing sophisticated
techniques to increase the key generation rate (which is currently 
limited to about $1$~Mbps~\cite{new1Mbps,highspeed1,highspeed2,highspeed3,highspeed4}) and
deal properly with various device imperfections of QKD implementations.
To extend the distance of QKD, the ideas of both
trusted and untrusted relay nodes have been studied.
There has also been much interest in the concept of ground to
satellite QKD. We will survey some of these recent efforts here.

{\it Multiplexing techniques:} Very recently, a field-test of 
a QKD system that multiplexes two quantum channels in the third telecom window using WDM has been performed~\cite{stableQKD}. The result is a 
very stable key generation rate from both channels over 30~days of operation without maintenance. 
This promising work confirms the possibility of using WDM 
techniques in QKD in order to increase its secure bit rate. Importantly, alternative results have also shown that quantum signals can 
be combined as well with 
strong conventional 
telecom traffics in the same fibre~\cite{Chapuran2009, Toshiba2012,Toshiba2014},
thus showing the feasibility of integrating QKD into existing fibre optical networks.
In~\cite{Chapuran2009}, for example, a QKD channel is located at $1310$nm, while classical channels 
use the third telecom window. A slight drawback of this solution, however, is the higher transmission loss of the fibre at 
$1310$nm, which limits the 
achievable QKD rate and distance. 
Alternatively, in~\cite{Toshiba2012,Toshiba2014} both the quantum and classical channels 
use wavelengths around $1550$nm. In so doing, \cite{Toshiba2012} achieves, for instance,
a secure key rate exceeding $1$ Mbps over 35km of fibre when the intensity of the classical signals is 
around $-18.6$~dBm. Remarkably, the same research group has shown  
that QKD is also possible in a high data laser power environment of around 0~dBm~\cite{Toshiba2014}. In this case,
the secret key rate is of the order of hundreds of Kbps over 25km of fibre.
On the other hand, it turns out that 
CV-QKD systems can also be quite robust against noise from strong telecom traffics due to multiplexing~\cite{Bing2010, DemBing2010}. 
This is so because the local oscillator serves as a ``mode selector''~\cite{mode selector} to suppress the noise.

{\it Development of the theory:}
The key generation rate can also be increased by developing better security analysis. 
A practical security proof must take into account the statistical fluctuations due to the finite data size. 
Therefore, the development of more sophisticated techniques for such analysis can result in higher key 
rates~\cite{Tsurumaru,Marcos,sec_prac}.
Also, one could include modifications in the protocol such as, for instance, the use of a biased basis choice.

{\it Extending the QKD coverage:} Up to this point, we have discussed different alternatives 
to integrate QKD into existing fibre optical networks and to
improve the key rate of the system. Another 
important parameter is the covered distance, which is typically limited to about $350$km (if one uses entanglement-based schemes). 
Of course, this upper limit could be improved if one employs 
ultra-low loss fibres~\cite{dpr9}. In general, one
simple solution to overcome this distance
limitation is to use trusted nodes, just like
in the QKD networks presented previously. Note, however, that to achieve secure communication over long distances
(say over $10,000$km) one needs many trusted nodes. Another possible solution is to use 
satellites, which could be employed both as trusted or untrusted nodes. In the former case, 
one can see the satellite as a trusted courier that can perform QKD as well as travel very fast around a certain orbit. This way, 
one could perform in the future QKD over the globe.
Indeed, a preliminary QKD experiment between ground and a hot-air balloon has been performed recently~\cite{sat2}; see also~\cite{sat}.  
This demonstration is illustrated in Fig.~1c.
It can be seen as a first step 
towards QKD between ground and 
Low Earth Orbit Satellites. Here, the development of accurate pointing techniques is one of 
the key technologies. Satellites could also be employed to build 
a QKD network with untrusted nodes by using, for example, 
measurement-device-independent (MDI) QKD~\cite{mdiQKD} (which will be discussed later in
the subsection on ``Counter-measures''),
where the parties on the ground send quantum signals
to the satellites that perform a joint measurement on the incoming signals. Also, one could place the source of an 
entanglement-based QKD protocol on a satellite and the receivers in the ground.
\begin{figure*}
\begin{center}
 \includegraphics[scale=0.324,angle=0]{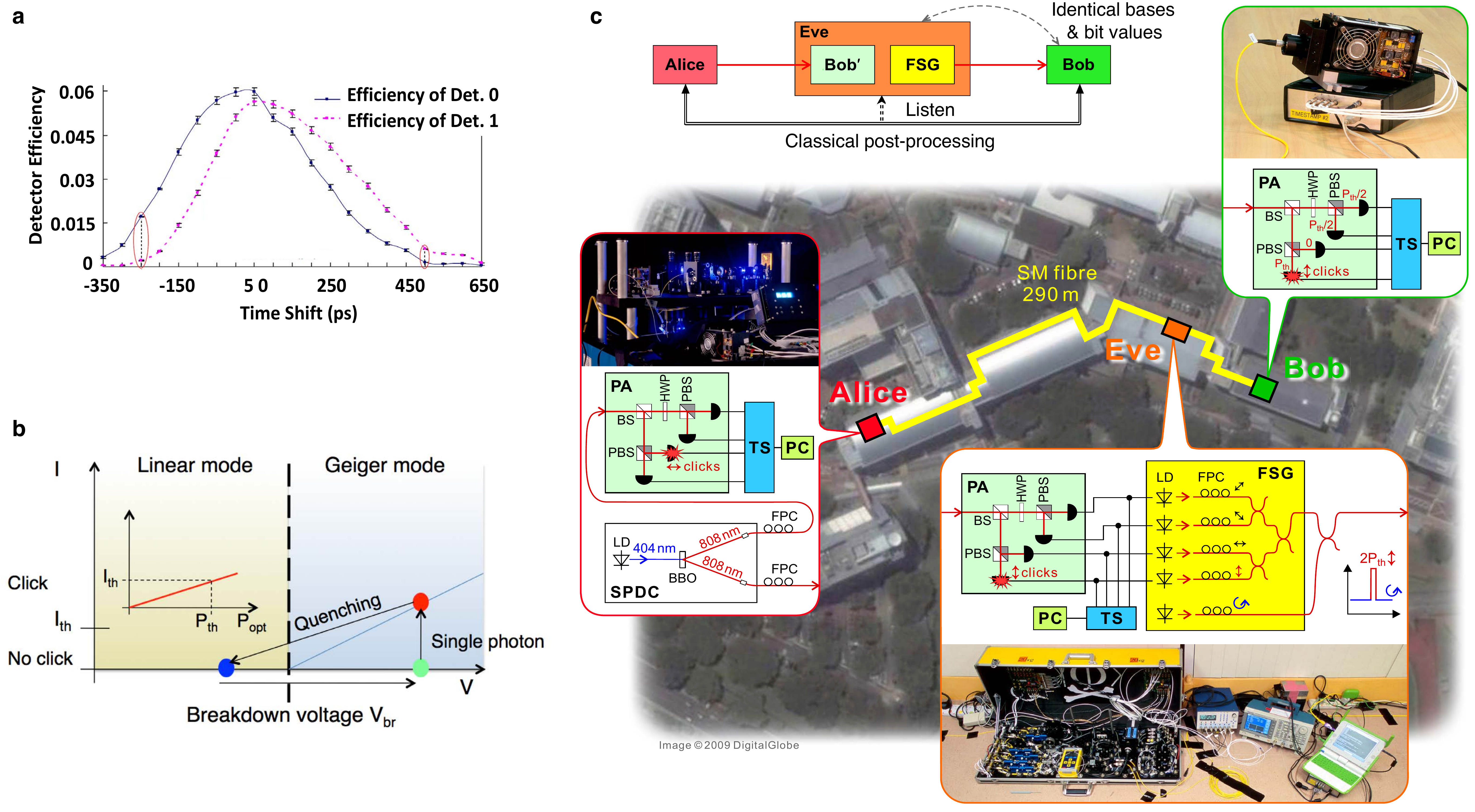}
 \end{center}
 \caption{
 {\bf $|$ Examples of quantum hacking.} {\bf a}, 
Experimentally measured detection efficiency mismatch between two detectors from a commercial QKD system 
versus time shifts~\cite{sidechannel}. This could be exploited by Eve to perform a time-shift attack~\cite{sidechannelrefs2a}, 
{\it i.e.}, she could shift the arrival time of each signal 
such that one detector has a much higher detection efficiency than the other. 
{\bf b},~Working principle of the detector blinding attack~\cite{sidechannelrefs2d}. By shining 
bright light into the detectors, Eve can make them leave the Geiger mode operation (used in QKD) and
enter into the linear mode operation. In so doing, she can control
which detector produces a ``click'' each given time and learn the entire secret key without being detected.
{\bf c}, Full-field implementation of a detector blinding attack on a running entanglement-based QKD set-up~\cite{sidechannelrefs2f}.
Figures from: {\bf a}, Ref.~\cite{sidechannel}; {\bf b}, Ref.~\cite{sidechannelrefs2d}, and {\bf c}, Ref.~\cite{sidechannelrefs2f}.
\label{fig5}}
\end{figure*} 

\quad

\noindent {\large\bf Quantum hacking and counter-measures} 

\noindent QKD can be proved secure in theory. However, are experimental implementations of QKD also secure? 
Security proofs rely on assumptions. Some of them are quite natural, such as, for instance, 
the validity of quantum mechanics. Other assumptions, however, are more severe, such as, 
for example, that Alice and Bob have an accurate and complete 
description of their physical apparatuses. Unfortunately, real-life realisations of QKD
often present imperfections and rarely conform to the 
theoretical models used to prove their security.  
As a result, there is a gap between the theory and the practice of QKD. Even though 
in principle QKD has been proven to be secure, practical systems may contain 
security loopholes, or so-called side-channels, which might be exploited by Eve 
to learn the distributed key without being detected. 

Indeed, this has been the case in recent attacks against certain 
commercial and research QKD 
set-ups~\cite{sidechannelrefs2a,sidechannel,time1,time2,sidechannelrefs2b,sidechannelrefs2d,shields1,reply_makarov,sidechannelrefs2f,sidechannelrefs2e,attackb,sidechannelrefs1c,attacka,newsc,newsc2,diamanti_sc},
where Eve employed some imperfections in the devices, specially in the single-photon detectors, to hack the system. 
But
one should not be overly alarmed by this fact at this stage, as current realisations of QKD are still in the battle-testing phase. 
Every time a new commercial cryptographic scheme is introduced, it is rather common for its first versions to contain some 
security flaws in the implementation. 
During the battle-testing period, these flaws are typically found and fixed. As a result, the systems become more and more secure. Also, 
it should be remarked that QKD is often 
combined with classical cryptography; for instance, by performing a bitwise XOR 
operation between 
a classical key and a key obtained with QKD.
In this sense, QKD can only improve the final security
of the whole system,
but not reduce it. 

\quad

\noindent {\bf Quantum hacking.}
What kind of imperfections can Eve exploit to hack a QKD system? In principle, QKD only secures the communication channel, so  
Eve may try to attack both the sources ({\it i.e.}, the preparation stage of the quantum signals) 
and the measurement device. A list of various existing attacks on QKD set-ups can be found in Table~\ref{table_hacking}.
\begin{table}
\begin{center}
    \begin{tabular}{ | c | c | c |}
    \hline \hline
    {\footnotesize \it Attack} & {\footnotesize \it Target component} & {\footnotesize \it Tested system} \\ \hline
    {\footnotesize Time-shift \cite{sidechannelrefs2a,sidechannel,time1,time2}} & {\footnotesize Detector} & {\footnotesize Commercial system} \\ \hline
    {\footnotesize Time-information \cite{sidechannelrefs2b}} & {\footnotesize Detector} & {\footnotesize Research system} \\ \hline
    {\footnotesize Detector-control \cite{sidechannelrefs2d,shields1,reply_makarov}} & {\footnotesize Detector} & {\footnotesize Commercial system} \\ \hline
    {\footnotesize Detector-control \cite{sidechannelrefs2f}} & {\footnotesize Detector} & {\footnotesize Research system} \\ \hline
    {\footnotesize Detector dead-time \cite{sidechannelrefs2e}} & {\footnotesize Detector} & {\footnotesize Research system} \\  \hline
    {\footnotesize Channel calibration \cite{attackb}} & {\footnotesize Detector} & {\footnotesize Commercial system} \\ \hline
    {\footnotesize Phase-remapping \cite{sidechannelrefs1c}} & {\footnotesize Phase modulator} & {\footnotesize Commercial system}  \\ \hline
    {\footnotesize Faraday-mirror \cite{attacka}} & {\footnotesize Faraday mirror} & {\footnotesize Theory} \\ \hline
    {\footnotesize Wavelength \cite{newsc}} & {\footnotesize Beam-splitter} & {\footnotesize Theory} \\ \hline
    {\footnotesize Phase information \cite{newsc2}} & {\footnotesize Source} & {\footnotesize Research system} \\ \hline
    {\footnotesize Device calibration \cite{diamanti_sc}} & {\footnotesize Local oscillator} & {\footnotesize Research system} \\ \hline
       \hline
    \end{tabular}
    \caption{Summary of various quantum hacking attacks against 
    certain commercial and research QKD set-ups.\label{table_hacking}}
\end{center}
\end{table}
The source is typically less likely to be a problem. This is so because Alice can prepare her quantum signals ({\it e.g.}, 
the polarisation state of phase-randomised 
WCPs) in a fully protected environment outside 
the influence of the eavesdropper. This can be achieved, for instance, using optical isolators. 
Also, Alice can 
experimentally verify the quantum states emitted. For this, she can employ, for example, random sampling techniques. 
Therefore, it is reasonable to expect that Alice can characterise her source. Fortunately, in this situation, 
it is usually relatively
easy to incorporate imperfections of Alice's state preparation process 
in the security proof~\cite{GLLP,kiyo13}. 

The problem with the measurement device of Bob is more subtle, 
as Eve is allowed to send in any signal she desires and, therefore, it is harder to protect Bob's set-up 
against any possible attack. Indeed, most quantum hacking strategies are directed at Bob's 
single-photon detectors~\cite{sidechannelrefs2a,sidechannel,time1,time2,sidechannelrefs2b,sidechannelrefs2d,shields1,reply_makarov,sidechannelrefs2f,sidechannelrefs2e,attackb}, 
which can be regarded as the Achilles' heel of QKD. For instance, Eve could exploit their detection efficiency mismatch~\cite{sidechannelrefs2a,sidechannel,time1,time2}. 
This is illustrated in Fig.~4a. 
However, the most important hacking attack so far against the detectors of the system is the so-called detector blinding attack~\cite{sidechannelrefs2d}. Here, Eve
shines bright light into the detectors to make
them enter into the so-called
linear mode operation, where they are not longer sensitive to single-photon pulses but only to strong light pulses~\cite{sidechannelrefs2d}. 
As a consequence, 
Eve can effectively fully control
which detector produces a ``click'' each given time, just by sending Bob
additional bright pulses. This way, Eve   
can learn the secret key completely. This is shown in 
Figs.~4b and~4c~\cite{sidechannelrefs2f}. Other possible imperfections that could be exploited 
include, for instance, the dead-time of the detectors~\cite{sidechannelrefs2e}. 

\quad

\begin{figure*}
\begin{center}
 \includegraphics[scale=0.32,angle=0]{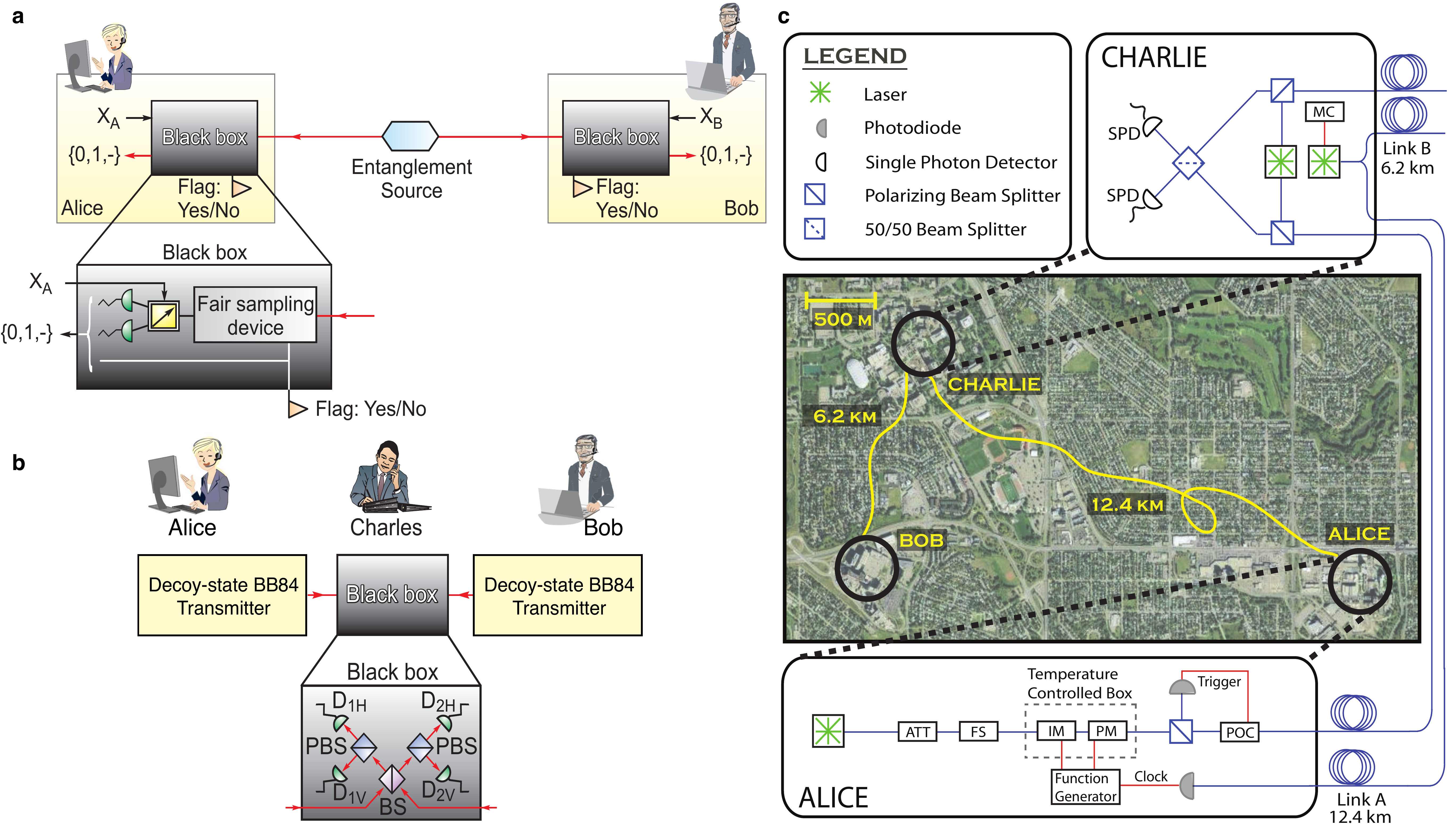}
 \end{center}
 \caption{
 {\bf $|$ Examples of counter-measures against quantum hacking.} {\bf a}, Schematic diagram of device-independent (DI) QKD~\cite{diqkdrefsa,diqkdrefse,diQKD1,diQKD2}. 
Alice and Bob can prove the security of the protocol based on the violation of an appropriate Bell inequality. 
To overcome the channel loss, the system can include a so-called 
fair sampling device~\cite{Gisin2010a,Gisin2010b}.  
In principle, DI-QKD can remove all side-channels in a QKD implementation.    
{\bf b}, 
Schematic representation of measurement-device-independent (MDI) QKD~\cite{mdiQKD}.
Alice and Bob prepare WCPs in different BB84 polarisation states and send them 
to an untrusted relay Charles, who is supposed to perform a Bell state measurement (BSM) that projects the incoming signals into a Bell state. 
MDI-QKD removes all detector side-channels, 
which can be regarded as the Achilles' heel of QKD. 
In comparison to DI-QKD, 
MDI-QKD has the advantage of being feasible with current technology. Indeed, 
proofs-of-principle demonstrations have been already done in~\cite{Rubenok2012a,Rubenok2012b}, 
and real QKD implementations have been realised in~\cite{Rubenok2012c,Zhiyuan:2013}.
{\bf c}, Field-test proof-of-principle demonstration of MDI-QKD realised 
in Calgary, Canada~\cite{Rubenok2012a}. Figures from: {\bf c}, Ref.~\cite{Rubenok2012a}.
\label{fig6}}
\end{figure*} 

\noindent {\bf Counter-measures.}
A natural solution to recover security in QKD implementations would be to develop mathematical models that perfectly match the behaviour of all QKD components and systems, 
and then incorporate this information into a new security proof. Unfortunately, while this is plausible in theory, it is hard to realise in practice (if not impossible), 
as QKD components are complex devices. Currently, there are three main alternative approaches. 

The first one is to use security patches. Fortunately, every time a security loophole is found, 
usually it is always quite easy to obtain a
suitable counter-measure~\cite{shields_counter,maka_com,shields_reply_com,avoid_sc_dps}. In so doing, one can guarantee 
security against known attacks. But note that the system might be defeated by hacking advances. 
This scenario is similar to most classical cryptographic techniques, as here one abandons the provable security 
model of QKD. 

The second approach is called device-independent (DI) QKD~\cite{diqkdrefsa,diqkdrefse,diQKD1,diQKD2}. 
See Fig.~5a.
Here, Alice and Bob treat their devices 
as two ``black boxes''. That is, they do not need to fully characterise their different elements. The security 
of DI-QKD relies on the 
violation of a Bell inequality, which certifies the presence of quantum correlations. Unfortunately, due to the detection 
efficiency loophole (which requires a detection efficiency around 80\% or higher), a loophole-free Bell test
is still unavailable. Indeed, 
the high decoupling and channel loss, together with the limited detection efficiency of current single-photon detectors,
renders DI-QKD highly impractical with current technology. Even if Alice and Bob try 
to compensate the channel loss by including a fair-sampling device, such as a qubit amplifier~\cite{Gisin2010a,Gisin2010b} 
or a quantum non-demolition 
measurement of the number of photons in a pulse, 
the 
resulting secret key rate of DI-QKD at practical distances is unfortunately very limited
({\it e.g.}, of order $10^{-10}$ bits per pulse)~\cite{Gisin2010a,Gisin2010b}. 
Of course, technology is improving
and DI-QKD might still prove viable in say 10 or 15 years.
In summary, the first approach to counter the quantum hacking problem is {\it ad hoc} 
whereas DI-QKD is impractical nowadays.

The third approach is MDI-QKD~\cite{mdiQKD},
which currently appears to be a possibly viable solution to the quantum hacking problem.
See Figs.~5b and~5c.
The main advantage of this approach is that it allows Alice and Bob to perform QKD with untrusted measurement devices, which can be even 
manufactured by Eve. 
In other words, MDI-QKD removes completely the weakest part of a QKD realisation, and offers an avenue to bridge the 
gap between theory and practice. The security of MDI-QKD is based on the idea of time reversal~\cite{biham,inamori}. 
Alice and Bob prepare quantum signals and send them to an untrusted relay, Charles/Eve, who is supposed to perform a 
Bell-state measurement on the signals received. The honesty of 
Charles can be verified by comparing a subset of the transmitted data. 
Most importantly, MDI-QKD
can be implemented with standard optical components
with low detection efficiency and highly lossy channels.
The key rate of MDI-QKD is many orders of magnitude higher than that of DI-QKD and the
experimental feasibility of MDI-QKD has been already demonstrated both in laboratories and via 
field-tests~\cite{Rubenok2012a,Rubenok2012b,Rubenok2012c,Zhiyuan:2013}. 
The key assumption in MDI-QKD is that Alice and Bob trust their sources. As noted earlier, this may
not be an unreasonable assumption because, when compared with single-photon detectors that receive unknown quantum states prepared by Eve, 
Alice and Bob have a much better chance to monitor their own preparation process carefully within their own laboratories. 
In fact, as mentioned before, source flaws can be taken
care off in security proofs~\cite{GLLP,kiyo13}.
A slight drawback of MDI-QKD is, however, its relatively low secret key rate when compared to the
decoy-state BB84 protocol. This is so because MDI-QKD requires
two-fold coincidence detector events which are suppressed due to
the low detection efficiency of standard InGaAs single-photon detectors. If one 
uses SNSPDs with 93\% detection efficiency as the ones described previously, then
such a disadvantage will disappear. Also, note that MDI-QKD could be used to build a QKD network 
with untrusted nodes, which would be desirable from a security standpoint.

\quad

\noindent{\large\bf Outlook}

\noindent To further extend the distance of secure quantum communication, there have been a lot of research activities in ``quantum repeaters"~\cite{rep1}, 
which allow entanglement to be swapped and distilled between different entangled pairs of photons.

If MDI-QKD is widely deployed in the future, then the frontier of
quantum hacking will shift towards attacking the source
(rather than the detectors). It will become important to
re-examine the various security assumptions used there
({\it e.g.}, single-mode assumption, perfect global phase randomisation
and no side-channels). The eternal welfare between the code-breakers and
code-makers will continue.

Owing to space limit, this article has focused on QKD. It should be noted that other applications of quantum cryptography such as 
quantum secret sharing, blind quantum computing and quantum coin flipping have also been proposed, whereas other protocols such as 
quantum bit commitment has been shown to be impossible without additional assumptions.

To conclude, we highlight the deep connections of quantum cryptography with other areas of physics as well as mathematics and technology. 
For instance, the loopholes in the security of practical QKD systems are closely related to the loopholes in the testing of Bell's inequalities in 
the foundations of quantum mechanics.  Quantum cryptography is also closely related to mathematics, information theory and statistics as it 
widely uses concepts in those subjects. Furthermore, quantum cryptography provides much impetus to technological developments in single-photon 
detectors, which can also be used to improve quantum metrology and sensing and contribute to the ultimate goal of the construction of large-scale quantum computers.

\bibliographystyle{apsrev}

\begin{thebibliography}{99}
\section*{References}

\bibitem{rsa} Rivest, R. L., Shamir A. \& Adleman, L. M.
A method of obtaining digital signatures and public-key cryptosystems.
{\it Commun. ACM} {\bf 21}, 120-126 (1978).

\bibitem{shor} Shor, P. W. 
Algorithms for quantum computation: discrete logarithms and factoring.
{\it in Proc. 35th Annual Symposium on Foundations of Computer Science}, ed. S. Goldwasser, 
(IEEE Computer Society Press, 1994), 124-134.
 
 \bibitem{bb84}
 Bennett, C. H. \& Brassard, G. 
 Quantum cryptography: public key distribution and coin tossing.
 {\it in Proc. IEEE International Conference on Computers, Systems, and Signal Processing}, 
 (IEEE Press, 1984), 175-179. 

\bibitem{vernam} Vernam, G. 
Cipher printing telegraph systems for secret wire and radio telegraphic communications.
{\it J. Am. Inst. Electr. Eng.} {\bf 45}, 109-115 (1926). 
 
 \bibitem{comp1} Ben-Or, M., Horodecki, M., Leung, D. W., Mayers, D. \& Oppenheim, J.
 The universal composable security of quantum key distribution.
{\it Theory of Cryptography: Second Theory of Cryptography Conference, TCC 2005}, 
Lecture Notes in Computer Science, 
ed. J. Kilian, (Springer Verlag, 2005),
{\bf 3378}, 386-406.
 
 \bibitem{comp2} Renner, R. \& Koenig, R.
 Universally composable privacy amplification against quantum adversaries. 
{\it Theory of Cryptography: Second Theory of Cryptography Conference, TCC 2005}, 
Lecture Notes in Computer Science, 
ed. J. Kilian, (Springer Verlag, 2005),
{\bf 3378}, 407-425.

\bibitem{stableQKD} Yoshino, K., Ochi, T., Fujiwara, M., Sasaki, M. \& Tajima, A.
Maintenance-free operation of WDM quantum key distribution system through a field fiber over 30 days.
{\it Opt. Express} {\bf 21}, 31395-31401 (2013).

\bibitem{free_QKD} Ursin, R. {\it et al.}
Entanglement-based quantum communication over 144Êkm.
{\it  Nature Phys.} {\bf 3}, 481-486 (2007).

\bibitem{net1} Elliott, C. {\it et al.} 
Current status of the DARPA Quantum Network. 
{\it Proc. SPIE} {\bf 5815}, 
(eds. Donkor, E. J., Pirich, A. R., and Brandt, H. E.), 138-149 (2005).

\bibitem{net2} Peev, M. {\it et al.}
The SECOQC quantum key distribution network in Vienna.
{\it New J. Phys.} {\bf 11}, 075001Ê(2009).

\bibitem{swissQ} Stucki, D. {\it et al.}
Long term performance of the SwissQuantum quantum key distribution network in a field environment.
{\it New J. Phys.} {\bf 13}, 123001 (2009).

\bibitem{China_net1} Chen, T.-Y. {\it et al.} 
Field test of a practical secure communication network with decoy-state quantum cryptography.
{\it Opt. Express} {\bf 17}, 6540-6549 (2009).

\bibitem{China_net2} Chen, T.-Y. {\it et al.} 
Metropolitan all-pass and inter-city quantum communication network.
{\it Opt. Express} {\bf 18}, 27217-27225 (2010).

\bibitem{ChineseQKD} Wang, S. {\it et al.}
Field test of wavelength-saving quantum key distribution network. 
{\it Opt. Lett.} {\bf 35}, 2454-2456 (2010).

\bibitem{demQKD} Sasaki, M. {\it et al.}
Field test of quantum key distribution in the Tokyo QKD network.
{\it Opt. Express} {\bf 19}, 10387-10409 (2011).

\bibitem{net3} Fr\"ohlich, B. {\it et al.}
A quantum access network.
{\it Nature} {\bf 501}, 69-72 (2013).

\bibitem{det1} Marsili, F. {\it et al.} 
Detecting single infrared photons with 93\% system efficiency.
{\it Nature Photon.} {\bf 7}, 210-214 (2013). 

\bibitem{High-efficient SNSPD21} Rosenberg, D., Kerman, A. J., Molnar, R. J. \& Dauler, E. A.
High-speed and high-efficiency superconducting nanowire single photon detector array. 
{\it Opt. Express} {\bf 21}, 1440-1447 (2013). 

\bibitem{High-efficient SNSPD3} Miki, S., Yamashita, T., Terai, H. \& Wang, Z.
High performance fiber-coupled NbTiN superconducting nanowire single photon detectors with Gifford-McMahon cryocooler. 
{\it Opt. Express} {\bf 21}, 10208-10214 (2013).

\bibitem{restelli13} Restelli, A., Bienfang, J. C. \& Migdall, A. L.
Single-photon detection efficiency up to 50\% at 1310nm with an InGaAs/InP avalanche diode gated at 1.25GHz.
{\it Appl. Phys. Lett.} {\bf 102}, 141104 (2013).

\bibitem{first} Bennett, C. H., Bessette, F., Brassard, G., Salvail, L. \& Smolin, J.
Experimental quantum cryptography.
{\it J. Cryptol.} {\bf 5}, 3-28 (1992). 

\bibitem{sat2} Wang, J.-Y. {\it et al.}
Direct and full-scale experimental verifications towards ground-satellite quantum key distribution.
{\it Nature Photon.} {\bf 7}, 387-393 (2013).

\bibitem{sat} Nauerth, S. {\it et al.}
Air-to-ground quantum communication.
{\it Nature Photon.} {\bf 7}, 382-386 (2013).

\bibitem{sec_prac} Lim, C. C. W., Curty, M., Walenta, N., Xu, F. \& Zbinden, H. 
Concise security bounds for practical decoy-state quantum key distribution.
{\it Phys. Rev. A} {\bf 89}, 022307 (2014).

\bibitem{PNS} Huttner, B., Imoto, N., Gisin, N., \& Mor, T.
Quantum cryptography with coherent states. 
{\it Phys. Rev. A} {\bf 51}, 1863-1869 (1995). 

\bibitem{GLLP}
Gottesman, D., Lo, H.-K., L\"utkenhaus, N. \& Preskill, J.
Security of quantum key distribution with imperfect devices. 
{\it Quant. Inf. Comp.} {\bf 5}, 325-360 (2004). 

\bibitem{Decoy1} Hwang, W.-Y. 
Quantum key distribution with high loss: toward global secure communication. 
{\it Phys. Rev. Lett.} {\bf 91}, 057901 (2003). 

\bibitem{Decoy2} Lo, H.-K., Ma, X. \& Chen, K.
Decoy state quantum key distribution. 
{\it Phys. Rev. Lett.} {\bf 94}, 230504 (2005).

\bibitem{Decoy3} Wang, X.-B. 
Beating the Photon-Number-Splitting attack in practical quantum cryptography. 
{\it Phys. Rev. Lett.} {\bf 94}, 230503 (2005).

\bibitem{Decoy4} Ma, X., Qi, B., Zhao, Y. \& Lo, H.-K.
Practical Decoy State for Quantum Key Distribution.
{\it Phys. Rev. A} {\bf 72}, 012326 (2005).

\bibitem{decoyexp1} Zhao, Y., Qi, B., Ma, X., Lo, H.-K. \& Qian, L. 
Experimental quantum key distribution with decoy states.
{\it Phys. Rev. Lett.} {\bf 96}, 070502 (2006).

\bibitem{decoyexp2} Peng, C.-Z. {\it et al.}
Experimental long-distance decoy-state quantum key distribution based on polarization encoding.
{\it Phys. Rev. Lett.} {\bf 98}, 010505 (2007).

\bibitem{decoyexp3} Rosenberg, D. {\it et al.}
Long-distance decoy-state quantum key distribution in optical fiber.
{\it Phys. Rev. Lett.} {\bf 98}, 010503 (2007).

\bibitem{decoyexp4} Schmitt-Manderbach, T. {\it et al.}
Experimental demonstration of free-space decoy-state quantum key distribution over 144 km.
{\it Phys. Rev. Lett.} {\bf 98}, 010504 (2007).

\bibitem{decoyexp5} Yuan, Z. L., Sharpe, A. W. \& Shields, A. J.
Unconditionally secure one-way quantum key distribution using decoy pulses.
{\it Appl. Phys. Lett.} {\bf 90}, 011118 (2007). 

\bibitem{decoyexp6} Liu, Y. {\it et al.}
Decoy-state quantum key distribution with polarized photons over 200km. 
{\it Opt. Express} {\bf 18}, 8587-8594 (2010).

\bibitem{Decoy-application} Wehner, S., Curty, M., Schaffner, C. \& Lo, H.-K.  
Implementation of two-party protocols in the noisy-storage model. 
{\it Phys. Rev. A} {\bf 81}, 052336 (2010).

\bibitem{hughesQKarD} 
Hughes, R. J. {\it et al.}
Network-Centric Quantum Communications with Application to Critical Infrastructure Protection.
{\it Preprint at http://lanl.arXiv.org/abs/1305.0305} (2013).

\bibitem{ekert91} Ekert, A. K. 
Quantum cryptography based on BellÕs theorem.
{\it Phys. Rev. Lett.} {\bf 67}, 661-663 (1991). 

\bibitem{ent0} Ma, X., Fung, C.-H. F. \& Lo, H.-K. 
Quantum key distribution with entangled photon sources. 
{\it Phys. Rev. A} {\bf 76}, 012307 (2007).

\bibitem{ent2} Treiber, A. {\it et al.}
Fully automated entanglement-based quantum cryptography system for telecom fiber networks.
{\it New J. Phys.} {\bf 11}, 045013 (2009). 

\bibitem{ent3} Poppe, A. {\it et al.}
Practical quantum key distribution with polarization-entangled photons.
{\it Opt. Express} {\bf 12}, 3865-3871 (2004). 

\bibitem{dpr1} Inoue, K., Waks, E. \& Yamamoto, Y. 
Differential phase shift quantum key distribution.
{\it Phys. Rev. Lett.} {\bf 89}, 037902 (2002). 

\bibitem{dpr5} Takesue, H. {\it et al.}   
Quantum key distribution over a 40 dB channel loss using superconducting single-photon detectors.
{\it Nature Photon.} {\bf 1}, 343-348 (2007).

\bibitem{dpr9} Stucki, D. {\it et al.}
High rate, long-distance quantum key distribution over 250km of ultra low loss fibres.
{\it New J. Phys.} {\bf 11}, 075003 (2009).

\bibitem{cv0} Grosshans, F. {\it et al.}
Quantum key distribution using gaussian-modulated coherent states.
{\it Nature} {\bf 421}, 238-241 (2003).

\bibitem{cv1} Qi, B., Huang, L.-L., Qian, L. \& Lo, H.-K. 
Experimental study on the Gaussian-modulated coherent-state quantum key distribution over standard telecommunication fibers.
{\it Phys. Rev. A} {\bf 76}, 052323 (2007).

\bibitem{cv4} Jouguet, P., Kunz-Jacques, S., Leverrier, A., Grangier, P. \& Diamanti, E.
Experimental demonstration of long-distance continuous-variable quantum key distribution.
{\it Nature Photon.} {\bf 7}, 378-381 (2013). 

\bibitem{det2} Yuan, Z. L., Kardynal, B. E., Sharpe, A. W. \& Shields, A. J. 
High speed single photon detection in the near-infrared.
{\it App. Phys. Lett.} {\bf 91}, 041114 (2007).

\bibitem{dixon09} Dixon, A. R. {\it et al.} 
Ultrashort dead time of photon-counting InGaAs avalanche photodiodes.
{\it Appl. Phys. Lett.} {\bf 94}, 231113 (2009).

\bibitem{namekata} Namekata, N., Sasamori, S. \& Inoue, S. 
800 MHz single-photon detection at 1550-nm using an InGaAs/InP avalanche photodiode operated with a sine wave gating.
{\it Opt. Express} {\bf 14}, 10043-10049 (2006).

\bibitem{liang} Liang, X.-L. {\it et al.}
Fully integrated InGaAs/InP single-photon detector module with gigahertz sine wave gating.
{\it Rev. Sci. Instrum.} {\bf 83}, 083111 (2012).

\bibitem{namekata2} Wu, Q.-L., Namekata. N. \& Inoue, S.
Sinusoidally Gated InGaAs Avalanche Photodiode with Direct Hold-Off Function for Efficient and Low-Noise Single-Photon Detection.
{\it Appl. Phys. Express} {\bf 6}, 062202 (2013).

\bibitem{Zhang09} Zhang, J., Thew, R., Barreiro, C. \& Zbinden, H. 
Practical fast gate rate InGaAs/InP single-photon avalanche photodiodes.
{\it Appl. Phys. Lett.} {\bf 95}, 091103 (2009).

\bibitem{Shibata} Shibata, H., Takesue, H., Honjo, T., Akazaki, T. \& Tokura, Y.
Single-photon detection using magnesium diboride superconducting nanowires. 
{\it Appl. Phys. Lett.} {\bf 97}, 212504 (2010).

\bibitem{random1} Pironio, S. {\it et al.}  
Random numbers certified by Bell's theorem.
{\it Nature} {\bf 464}, 1021-1024 (2010).

\bibitem{random2b} Williams, C. R. S., Salevan, J. C., Li, X., Roy, R. \& Murphy, T. E.
Fast physical random number generator using amplified spontaneous emission. 
{\it Opt. Express} {\bf 18}, 23584-23597 (2010).

\bibitem{random3} Jofre, M. {\it et al.}  
True random numbers from amplified quantum vacuum.
{\it Opt. Express} {\bf 19}, 20665-20672 (2011).

\bibitem{random3b} Abell\'an, C. {\it et al.} 
Ultra-fast quantum randomness generation by accelerated phase diffusion in a pulsed laser diode.
{\it Opt. Express} {\bf 22}, 1645-1654 (2014).

\bibitem{random4} Qi, B., Chi, Y.-M., Lo, H.-K. \& Qian, L.
Experimental demonstration of a high speed quantum random number generation scheme based on measuring phase noise of a single mode laser.
{\it Opt. Lett.} {\bf 35}, 312-314 (2010).

 \bibitem{Townsend} Choi, I., Young, R. J. \& Townsend, P. D.
Quantum key distribution on a 10Gb/s WDM-PON. 
{\it Opt. Express} {\bf 18}, 9600-9612 (2010).

\bibitem{Toshiba2012} Patel, K. A. {\it et al.}
Coexistence of high-bit-rate quantum key distribution and data on optical fiber. 
{\it Phys. Rev. X} {\bf 2}, 041010 (2012).

\bibitem{Chapuran2009} Chapuran, T. E. {\it et al.} 
Optical networking for quantum key distribution and quantum communications.
{\it New J. Phys.} {\bf 11}, 105001 (2009). 

\bibitem{Toshiba2014} Patel, K. A. {\it et al.}
Quantum key distribution for 10 Gb/s dense wavelength division multiplexing networks.
{\it Appl. Phys. Lett.} {\bf 104}, 051123 (2014).

\bibitem{new1Mbps} Dixon, A. R., Yuan, Z. L., Dynes, J. F., Sharpe, A. W. \& Shields, A. J.
Gigahertz decoy quantum key distribution with 1 Mbit/s secure key rate.
{\it Opt. Express} {\bf 16}, 18790-18979 (2008).

\bibitem{highspeed1} Zhang, Q. {\it et al.}
Megabits Secure Key Rate Quantum Key Distribution.
{\it New J. Phys.} {\bf 11}, 045010 (2009).

\bibitem{highspeed2} Dixon, A. R., Yuan, Z. L., Dynes, J. F., Sharpe, A. W. \& Shields, A. J.
Continuous operation of high bit rate quantum key distribution.
{\it Appl. Phys. Lett.} {\bf 96}, 161102 (2010).

\bibitem{highspeed3} Tanaka, A. {\it et al.}
High-Speed Quantum Key Distribution System for 1-Mbps Real-Time Key Generation.
{\it IEEE J. Quant. Electron.} {\bf 48}, 542-550 (2012).

\bibitem{highspeed4} Walenta, N. {\it et al.}
1 Mbps coherent one-way QKD with dense wavelength division multiplexing and hardware key distillation, 
abstract, 
 2nd Annual Conference on Quantum Cryptography, Singapore, (2012); available at 
 http://2012.qcrypt.net/docs/extended-abstracts/qcrypt2012$\_$submission$\_$35.pdf.
 
\bibitem{Bing2010} Qi, B., Zhu, W., Qian, L. \& Lo, H.-K.
Feasibility of quantum key distribution through dense wavelength division multiplexing network. 
{\it New J. Phys.} {\bf 12}, 103042 (2010).

\bibitem{DemBing2010} Jouguet, P. {\it et al.} 
Experimental demonstration of the coexistence of continuous-variable quantum key distribution with an intense DWDM classical channel, 
abstract, 3rd Annual Conference on Quantum Cryptography, Waterloo, Canada, (2013); available at 
http://2013.qcrypt.net/contributions/Jouguet-abstract.pdf.

\bibitem{mode selector} Raymer, M. G., Cooper, J., Carmichael, H. J., Beck M. \& Smithey, D. T.
Ultrafast measurement of optical-field statistics by dc-balanced homodyne detection. 
{\it J. Opt. Soc. Am. B} {\bf 12}, 1801-1812 (1995).

\bibitem{Tsurumaru} Hayashi, M. \& Tsurumaru, T. 
Concise and tight security analysis of the Bennett-Brassard 1984 protocol with finite key lengths. 
{\it New J. Phys.} {\bf 14}, 093014 (2012).

\bibitem{Marcos} Curty, M. {\it et al.} 
Finite-key analysis for measurement-device-independent quantum key distribution. 
{\it Nature Commun.} \textbf{5}, 3732 (2014).

\bibitem{mdiQKD} Lo, H.-K., Curty, M. \& Qi, B. 
Measurement-device-independent quantum key distribution.
{\it Phys. Rev. Lett.} {\bf 108}, 130503 (2012).

\bibitem{sidechannelrefs2a}
 Qi, B., Fung, C.-H. F., Lo, H.-K. \& Ma, X. 
 Time-shift attack in practical quantum cryptosystems.
 {\it Quant. Inf. Comp.} {\bf 7}, 73-82 (2007).

\bibitem{sidechannel} 
Zhao, Y., Fung, C.-H. F., Qi, B., Chen, C. \& Lo, H.-K.
Quantum hacking: experimental demonstration of time-shift attack against practical quantum-key-distribution systems.
{\it Phys. Rev. A} {\bf 78}, 042333 (2008).

\bibitem{time1} Makarov, V., Anisimov, A. \& Skaar, J. 
Effects of detector efficiency mismatch on security of quantum cryptosystems.
{\it Phys. Rev. A} {\bf 74}, 022313 (2006).

\bibitem{time2} Makarov, V., Anisimov, A. \& Skaar, J. 
Erratum: Effects of detector efficiency mismatch on security of quantum cryptosystems [Phys. Rev. A 74, 022313 (2006)].
{\it Phys. Rev. A} {\bf 78}, 019905 (2008).

\bibitem{sidechannelrefs2b}
 Lamas-Linares, A. \& Kurtsiefer, C. 
 Breaking a quantum key distribution system through a timing side channel.
{\it Opt. Express} \textbf{15}, 9388-9393 (2007).

\bibitem{sidechannelrefs2d}
 Lydersen, L. {\em et al.}
 Hacking commercial quantum cryptography systems by tailored bright illumination. 
{\it Nature Photon.} \textbf{4}, 686-689 (2010).

\bibitem{shields1} Yuan, Z. L., Dynes, J. F. \& Shields, A. J.
Avoiding the blinding attack in QKD. 
{\it Nature Photon.} \textbf{4}, 800-801 (2010).

\bibitem{reply_makarov} Lydersen, L. {\em et al.}
Reply to ``Avoiding the blinding attack in QKD''.
{\it Nature Photon.} \textbf{4}, 801 (2010).

\bibitem{sidechannelrefs2f}
 Gerhardt, I. {\em et al.}
 Full-field implementation of a perfect eavesdropper on a quantum cryptography system.
 {\it Nature Commun.} \textbf{2}, 349 (2011).
 
\bibitem{sidechannelrefs2e}
 Weier, H. {\em et al.}
Quantum eavesdropping without interception: an attack exploiting the dead time of single-photon detectors.
{\it New J. Phys.} \textbf{13}, 073024 (2011). 

\bibitem{attackb}
Jain, N. {\em et al.}
Device calibration impacts security of quantum key distribution.
{\it Phys. Rev. Lett.} {\bf 107}, 110501 (2011).

\bibitem{sidechannelrefs1c}
Xu, F., Qi, B. \& Lo, H.-K. 
Experimental demonstration of phase-remapping attack in a practical quantum key distribution system.
{\it New J. Phys.} \textbf{12}, 113026 (2010).

\bibitem{attacka}
Sun, S.-H., Jiang, M.-S. \& Liang, L.-M.
Passive Faraday-mirror attack in a practical two-way quantum-key-distribution system.
{\it Phys. Rev. A} {\bf 83}, 062331 (2011).

\bibitem{newsc} Huang, J.-Z. {\em et al.}
Quantum hacking on continuous-variable quantum key distribution system using a wavelength attack.
{\it Phys. Rev. A} {\bf 87}, 062329 (2013).

\bibitem{newsc2} Tang, Y.-L. {\em et al.}
Source attack of decoy-state quantum key distribution using phase information.
{\it Phys. Rev. A} {\bf 88}, 022308 (2013).

\bibitem{diamanti_sc} Jouguet, P., Kunz-Jacques, S. \& Diamanti, E. 
Preventing calibration attacks on the local oscillator in continuous-variable quantum key distribution.
{\it Phys. Rev. A} {\bf 87}, 062313 (2013).

\bibitem{kiyo13}
Tamaki, K., Curty, M., Kato, G., Lo, H.-K. \& Azuma, K. 
Loss-tolerant quantum cryptography with imperfect sources.
{\it Preprint at http://lanl.arXiv.org/abs/1312.3514} (2013).

\bibitem{shields_counter} Yuan, Z. L., Dynes, J. F. \& Shields, A. J.
Resilience of gated avalanche photodiodes against bright illumination attacks in quantum cryptography.
{\it Appl. Phys. Lett.} {\bf 98}, 231104 (2011).

\bibitem{maka_com} Lydersen, L., Makarov, V. \& Skaar, J.
Comment on ``Resilience of gated avalanche photodiodes against bright illumination attacks in quantum cryptography''.
{\it Appl. Phys. Lett.} {\bf 99}, 196101 (2011).

\bibitem{shields_reply_com} Yuan, Z. L., Dynes, J. F. \& Shields, A. J.
Response to ``Comment on `Resilience of gated avalanche photodiodes against bright illumination attacks in quantum cryptography'''
{\it Appl. Phys. Lett.} {\bf 99}, 196102 (2011).

\bibitem{avoid_sc_dps} Honjo, T. {\em et al.}
Countermeasure against tailored bright illumination attack for DPS-QKD.
{\it Opt. Express} {\bf 21}, 266-2673 (2013).

\bibitem{diqkdrefsa}
Mayers, D. \& Yao, A. C.-C. 
Quantum cryptography with imperfect apparatus.
{\it in Proceedings of the 39th Annual Symposium on Foundations of Computer Science},
503-509  (IEEE Computer Society Press, 1998).

\bibitem{diqkdrefse}
Masanes, L., Pironio, S. \& Ac\'{i}n, A. 
Secure device-independent quantum key distribution with causally independent measurement devices.
{\it Nature Commun.} {\textbf 2}, 238 (2011).

\bibitem{diQKD1} Reichardt, B. W., Unger, F. \&  Vazirani, U. 
Classical command of quantum systems.
{\it Nature} {\bf 496}, 456-460 (2013).

\bibitem{diQKD2} Vazirani, U. \& Vidick, T. 
Fully device independent quantum key distribution. 
{\it Preprint at http://lanl.arXiv.org/abs/1210.1810} (2012).

\bibitem{Gisin2010a}
Gisin, N., Pironio, S. \& Sangouard, N. 
Proposal for implementing device-independent quantum key distribution based on a heralded qubit amplifier.
{\it Phys. Rev. Lett.} {\bf 105}, 070501 (2010).

\bibitem{Gisin2010b}
Curty, M. \& Moroder, T. 
Heralded-qubit amplifiers for practical device-independent quantum key distribution.
{\it Phys. Rev. A.} {\bf 84}, 010304(R) (2011).
 
\bibitem{Rubenok2012a}
Rubenok, A., Slater, J. A., Chan, P., Lucio-Martinez, I. \& Tittel, W.
Real-world two-photon interference and proof-of-principle quantum key distribution immune to detector attacks.
{\it Phys. Rev. Lett.} {\bf 111}, 130501 (2013).

\bibitem{Rubenok2012b}
Ferreira da Silva, T. {\em et al.}
Proof-of-principle demonstration of measurement device independent QKD using polarization qubits.
{\it Phys. Rev. A} {\bf 88}, 052303 (2013).

\bibitem{Rubenok2012c}
Liu, Y. {\em et al.}
Experimental measurement-device-independent quantum key distribution. 
{\it Phys. Rev. Lett.} {\bf 111}, 130502 (2013).

\bibitem{Zhiyuan:2013}
Tang, Z. {\em et al.}  
Experimental demonstration of polarization encoding measurement-device-independent quantum key distribution.
{\it Preprint at http://lanl.arXiv.org/abs/1306.6134} (2013). 

\bibitem{biham}
Biham, E., Huttner, B. \& Mor, T. 
Quantum cryptographic network based on quantum memories.
{\it Phys. Rev. A} {\bf 54}, 2651-2658 (1996).

\bibitem{inamori}
Inamori, H. 
Security of practical time-reversed EPR quantum key distribution.
{\it Algorithmica} {\bf 34}, 340-365 (2002).


\bibitem{rep1} Briegel, H. J., D\"ur, W., Cirac, J. I. \& Zoller, P. 
Quantum repeaters: the role of imperfect local operations in quantum communication.
{\it Phys. Rev. Lett.} {\bf 81}, 5932-5935 (1998).

\end{thebibliography}

\section*{Acknowledgments}

The authors thank 
K. Azuma, C. H. Bennett, M. Fujiwara, G. Kato, N. Matsuda, N. Namekata, T. Ochi, B. Qi, L. Qian, M. Sasaki, H. Shibata, H. Takesue, F. Xu, K. Yoshino, 
Q. Zhang, Y. Zhao and the anonymous referees for their valuable comments and suggestions.
We specially thank C. H. Bennett and R. J. Hughes for allowing us to use
a photo of the first experimental demonstration of QKD
and a photo of the first-generation, modularly-integrated
QKarD respectively in this Review Article.
We acknowledge support from the European Regional Development Fund (ERDF), the Galician Regional Government (projects CN2012/279 and CN 2012/260, 
``Consolidation of Research Units: AtlantTIC''), NSERC, the CRC program, 
the Connaught Innovation Award,
the project
``Secure photonic network technology'' as part of ``The project UQCC'' by the National Institute of Information and Communications Technology (NICT) of Japan, and from the Japan Society for the Promotion of Science (JSPS) through its Funding Program for World-Leading Innovative R\&D on Science and Technology (FIRST Program).


\end{document}